\documentclass[12pt]{iopart}
\usepackage{amsfonts}
\usepackage{graphicx}

\begin{document}

\title[Collisions and evaluation of spin-dependent interactions]
{Repeating head-on collisions in an optical trap and the evaluation
of spin-dependent interactions among neutral particles}

\author{Z B Li, Z F Chen, Y Z He, C G Bao\footnote{The corresponding author}}

\address{State Key Laboratory of Optoelectronic Materials and Technologies, and
Department of Physics, Sun Yat-Sen University, Guangzhou, 510275,
P.R. China}

\ead{stsbcg@mail.sysu.edu.cn}

\begin{abstract}
A dynamic process of repeating collisions of a pair of trapped
neutral particles with weak spin-dependent interaction is designed
and studied. Related theoretical derivation and numerical
calculation have been performed to study the inherent
coordinate-spin and momentum-spin correlation. Due to the repeating
collisions the effect of the weak interaction can be accumulated and
enlarged, and therefore can be eventually detected. Numerical
results suggest that the Cr-Cr interaction, which has not yet been
completely clear, could be thereby determined. The design can be in
general used to determine various interactions among neutral atoms
and molecules, in particular for the determination of very weak
forces.
\end{abstract}

\pacs{34.90.+q, 34.50.Cx, 03.75.Mn}

\maketitle

\section{Introduction}

It is well known that the study of scattering is an important way to
understand the interactions among particles. Experimentally, the
incident particles with certain momentum would hit a target to
initiate the scattering. The accuracy of experimental data depends
on a precise control of bombarding energy. For charged incident
particles, the initial momenta are usually imposed by
electromagnetic forces which can be precisely tuned. For neutral
incident particles, the initial momenta are in general difficult to
control precisely. Furthermore, the interactions among neutral atoms
or molecules are in general weak (say, the van der Waals force).
Therefore, the precise determination of these interactions is
difficult. Consequently, these interactions are usually determined
in an indirect way (say, comparing related experimental data of
spectroscopy and/or thermodynamics of a many-body system with
theoretical results based on a model with a given set of
parameters). In this way the associated theoretical calculations are
usually complicated, and uncertainty might exist.

On the other hand, the trapping of neutral atoms has been realized
via optical traps since 1998 \cite{r_SDM1998,r_SJ1998}. This
techniques is in progress and fewer atoms can be trapped recently
\cite{r_AM2005}. It might open a new way for studying the
interactions. In this paper, an idea is proposed and related
theoretical calculation is performed to show how the scattering with
precisely controllable initial status can be realized in a trap. It
turns out that, as we shall see, the collisions would occur
repeatedly. Thereby the effect of each individual collision can be
accumulated and therefore enlarged. This might lead to a better
understanding of very weak interactions among neutral particles.

Since the pioneer experiment by Greismaier, et. al. \cite{r_GA2005}, the
Bose-Einstein condensations of atoms with a larger spin (say, $^{52}$Cr)
become a hot topic. These condensates are a new kind of matter aggregation
having the magnetic dipole-dipole interaction $V_{dd}$ more than twenty
times stronger than that of the alkalis family.
\begin{equation}
 V_{dd}
 =\frac{C_{d}}{r^{3}}
  \Big[\mathbf{F}_1 \cdot \mathbf{F}_2
   -3\frac{(\mathbf{F}_1 \cdot \mathbf{r})
           (\mathbf{F}_2 \cdot \mathbf{r})}
          {r^2}\Big],
 \label{e01_Vdd}
\end{equation}
where the strength $C_d=\mu_0 \mu_B^2 g_F^2 /(4\pi)$ with $\mu_0$
being the magnetic permeability of vacuum, $\mu_B$ the Bohr
magneton, and $g_F$ the Land\'{e} $g$ factor, $\mathbf{F}_i$ the
operator of the spin of the $i$-th atom, and
$\mathbf{r}=\mathbf{r}_2-\mathbf{r}_1$. Consequently, the spatial
and spin degrees of freedom are coupled so that the conversion of
spin angular momentum into orbital angular momentum can be realized.
Thereby new physical phenomena (say, rotonlike behavior) might
appear \cite{r_DD2003,r_SAN2003}. In addition to the long range
$V_{dd}$, the short range interaction is also spin-dependent and,
for low-energy systems, can be in general written as $V_{\delta} =
\delta (\mathbf{r}_1 - \mathbf{r}_2)\sum_S g_S \mathfrak{P}^S$,
where $S$ is the total spin, $g_S$ is the strength related to the
$s$-wave scattering length of the $S$-spin channel, and
$\mathfrak{P}^S$ the projector of the $S$-channel. $g_S$ is nonzero
only if $S$ is even. $g_2$, $g_4$, and $g_6$ are known while $g_0$
has not yet \cite{r_WJ2005,r_SJ2005,r_GA2006,r_DRB2006}. However,
many features of the condensate depend strongly on $g_0$ (say, the
phase-diagrams \cite{r_DRB2006,r_MH2007,r_US2008,r_VP2007} and the
spin-evolutions \cite{r_SL2006}). Therefore, the determination of
$g_0$ is important for a thorough and clear description of this
condensate. As an application of our idea of repeating collisions in
a trap, the $^{52}$Cr atoms have been chosen as an example to see to
what extent the interaction can be thereby clarified.

In the beginning, two narrow and deep potentials $\frac{1}{2}m\omega
_a^2|\mathbf{r\pm a}|^2$ are preset at $\mathbf{\pm a}$ (say, two
magnetic traps), where $\mathbf{a}$ is lying along the positive
$Z$-axis. Each potential contains a Cr atom in the ground state of
the parabolic confinement. Both atoms are polarized but in reverse
directions. The upper (lower) atom has spin-component $\mu =3\
(-3)$. Thus the magnetization $M_S$ of the system is zero, and the
two atoms are localized. Instantly, the two preset potentials are
replaced by a broader potential $\frac{1}{2} m \omega^2 r^2$ (say,
an optical trap) centering at the origin. Then, the previously
localized atoms begin to evolve. Since $M_S=0$, the $S=0$ component
must be included and will be affected by $g_0$. Therefore, by
observing the evolution, the knowledge on $g_0$ might be extracted.

\section{Hamiltonian and initial state}

Introduce $\mathbf{R}=(\mathbf{r}_1+\mathbf{r}_2)/2$ and
$\mathbf{r}=\mathbf{r}_2-\mathbf{r}_1$ for the c.m. and relative
motions. Introduce $\hbar \omega$ and $\sqrt{\hbar /m\omega }$ as
the units of energy and length, respectively. The symmetrized and
normalized initial state
\begin{eqnarray}
 \Psi_{I}
 &=&\frac{1+P_{12}}{\sqrt{2}}
  \Big[\Big(\frac{2\alpha}{\pi}\Big)^{3/4}
   e^{-\alpha R^2}\Big] \nonumber \\
 &&\cdot
  \Big[\Big(\frac{\alpha }{2\pi}\Big)^{3/4}
   e^{-\alpha (r^2/4 + a^2 + ra\cos\theta)}\Big]
  \chi_3(1)
  \chi_{-3}(2),
 \label{e02_PsiI}
\end{eqnarray}
where $P_{12}$ denotes an interchange of 1 and 2,
$\alpha=\omega_a/\omega$, $\theta$ is the angle between $\mathbf{r}$
and the $Z$-axis, $a=|\mathbf{a}|$, and $\chi_{\mu}$ is a spin-state
of an atom with component $\mu$. Then the evolution is governed by
the Hamiltonian
\begin{equation}
 H_{evol} = H_R + H_r +V_{12},
 \label{e03_Hevol}
\end{equation}
where $H_{R} \equiv -\frac{1}{4} \nabla_R^2 + R^2$ and $H_r \equiv
-\nabla_r^2 + \frac{1}{4} r^2$, and $V_{12}=V_{\delta }+V_{dd}$. The
eigenstates of $H_R$ and $H_r$, denoted as
$\bar{\phi}_{NL}(R)Y_{LM}(\hat{R})$ and
$\phi_{nl}(r)Y_{lm_{l}}(\hat{r})$, are just the harmonic oscillator
states. The eigenstates of $H_r + V_{12}$ can be expanded in terms
of basis functions as
\begin{equation}
 \psi_i^J=\sum_{\gamma}C_{i\gamma}^J\phi_{nl}(r)(lS)_{J},
 \label{e04_psiiJ}
\end{equation}
where $\gamma$ represents the set $n$, $l$, and $S$, $(lS)_J$
denotes the coupling of $l$ and $S$ into the total angular momentum
$J$, $i$ is just an index of the $J$-series. Due to $V_{dd}$, $l$
and $S$ are not conserved, but $J$ is. Due to the boson statistics,
$l+S$ must be even.

It turns out that $V_{12}$ is rather weak in our case. Consequently,
each eigenstate of $H_r+V_{12}$ is close to an eigenstate of $H_r$.
This fact leads to a great reduction of necessary basis functions in
the expansion. When a set of basis functions has been chosen, the
associated matrix elements of $H_r+V_{12}$ can be derived as shown
in the appendix. Carrying out the diagonalization, the coefficients
$C_{i\gamma}^J$ and the corresponding eigenenergy $E_i^J$ can be
obtained. In terms of $\bar{\phi}_{NL}$ and $\psi_i^J$, the initial
state can be rewritten as
\begin{equation}
 \Psi_I=\sum_N B_N \bar{\phi}_{N,0}(R) Y_{00}(\hat{R})
        \sum_{J,i}b_i^J \psi_i^J.
 \label{e05_PsiI}
\end{equation}

Making use of Eq.~(\ref{e04_psiiJ}), and equating (\ref{e05_PsiI})
and (\ref{e02_PsiI}), it is straight forward to obtain the
coefficients $B_N$ and $b_i^J$.

\section{Time-dependent density and the repeating collisions}

The time-dependent solution for the evolution is
\begin{eqnarray}
 \Psi(t)
 &=&e^{-iH_{evol}\tau}
  \Psi_I \nonumber \\
 &=&\frac{1}{\sqrt{4\pi}}
  \sum_N
  B_N
  e^{-i(2N+\frac{3}{2})\tau}
  \bar{\phi}_{N,0}(R) \nonumber \\
 &&\cdot
  \sum_{J,i}
  b_i^J
  e^{-i E_i^J \tau}
  \psi_i^J,
 \label{e06_Psit}
\end{eqnarray}
where $\tau =\omega t$.

From Eq.~(\ref{e06_Psit}), all the information on the evolution can
be extracted. The main feature of the evolution is the occurrence of
repeated collisions, as we shall see, and the effect of interaction
on each collision can be accumulated and therefore enlarged. Thereby
the strength $g_0$ can be evaluated.

We firstly extract the time-dependent density from $\Psi(t)$ as
\begin{eqnarray}
 \rho(r,\theta,t)
 &\equiv& 2\pi r^2
          \int
          d\mathbf{R}
          \Psi^*(t) \Psi(t) \nonumber \\
 &=& 2\pi r^2
     \sum_{Ji\gamma,J'i'\gamma'}
     \delta_{S',S}
     \cos[(E_{i'}^{J'}-E_i^J)\tau] \nonumber \\
 &&\cdot
     b_{i'}^{J'}
     C_{i'\gamma'}^{J'}
     b_i^J
     C_{i\gamma}^J
     \sum_{M_S}
     C_{l',-M_S;\ SM_S}^{J',0}
     C_{l,-M_S;\ SM_S}^{J,0} \nonumber \\
 &&\cdot
     \phi_{n'l'}(r)
     \phi_{nl}(r)
     |Y_{l',-M_S}|
     |Y_{l,-M_S}|,
 \label{e07_rhorqt}
\end{eqnarray}
where the Clebsch-Gordan coefficients have been introduced. It satisfies
\begin{equation}
 1=\int dr \sin\theta \ d\theta \ \rho(r,\theta,t).
 \label{e08_1}
\end{equation}

\begin{figure}[htbp]
 \centering
 \resizebox{0.95\columnwidth}{!}{\includegraphics{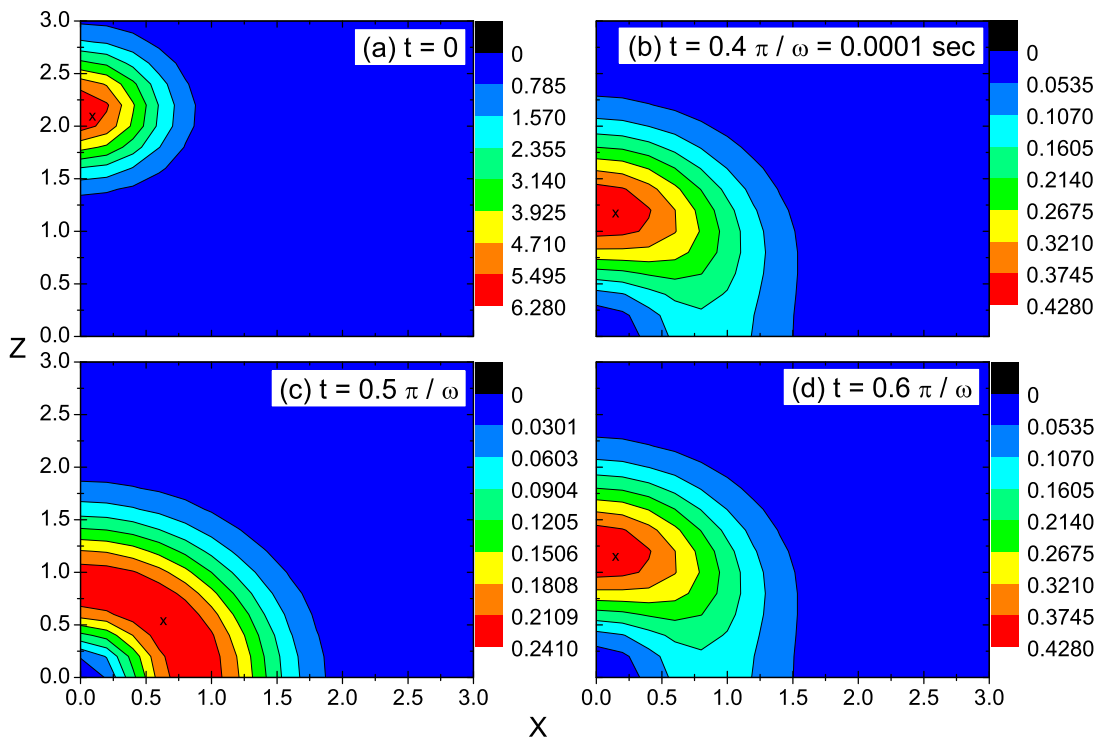}}
 \caption{(Color online) $\rho (r,\protect\theta,t)$ plotted in the
$z$-$x$ plane, where $z$ and $x$ are the components of
$\mathbf{r}/2$. $t$ is given at $4$ values in the interval ($0$ to
$0.6\pi/\omega=0.00015\sec$). In the panel, the area with the
largest $\rho(r,\theta,t)$ is marked by a $\times$. The parameters
are given as $\omega=2000\times 2\pi$, $\alpha=1.5$,
$|\mathbf{a}|=2\sqrt{\hbar/m\omega }=311~nm$, and $g_0=-g_6/2$.}
 \label{revfig1}
\end{figure}

When the parameters are given as $\omega=2000\times 2\pi$,
$\alpha=1.5$, $a=2$, and $g_0=-g_6/2$, the variation of $\rho$ in
the earliest stage of evolution is shown in Fig.~\ref{revfig1}.
Since $\rho$ does not depend on the azimuthal angle and
$\rho(r,\pi-\theta,t)=\rho(r,\theta,t)$, it is sufficient to be
plotted only on a quarter of $z$-$x$ plane, where $z$ and $x$ are
the components of $\mathbf{r}/2$. Since the c.m. is always
distributed close to the origin, we have roughly
$\mathbf{r}_{2}\mathbf{\simeq r}/2$ and
$\mathbf{r}_1\simeq-\mathbf{r}/2$. Thus the distribution of an atom
can be understood from Fig.~\ref{revfig1}. When the evolution
begins, the two atoms located at opposite ends of the broad
potential collide straightly with each other (see \ref{revfig1}a and
\ref{revfig1}b). When $t\approx\pi/2\omega$ (\ref{revfig1}c), the
two atoms keep close to each other with a distance $\sim 1.6$, and
they are both distributed around the center. Afterward, the atoms
begin to separate (\ref{revfig1}d is very similar to
\ref{revfig1}b). When $t=\pi/\omega$, the profile (not yet shown) is
very similar to \ref{revfig1}a. Thus the first round of head-on
collision has been completed, and the second round will begin
successively. If we remove $V_{12}$ from $H_{evol}$, the factor
$(E_{i'}^{J'}-E_i^J)$ in Eq.~(\ref{e07_rhorqt}) would become an
integral multiple of $\hbar\omega$, thereby the above process would
be exactly periodic with the period $2\pi/\omega$. In fact, the
collision as shown in Fig.~\ref{revfig1} is essentially determined
by $H_r$. In the early stage, $V_{12}$ causes only a very small
perturbation. However, as we shall see, the effect of each collision
can be accumulated when the time goes on.
\begin{figure}[tbph]
 \centering
 \resizebox{0.95\columnwidth}{!}{\includegraphics{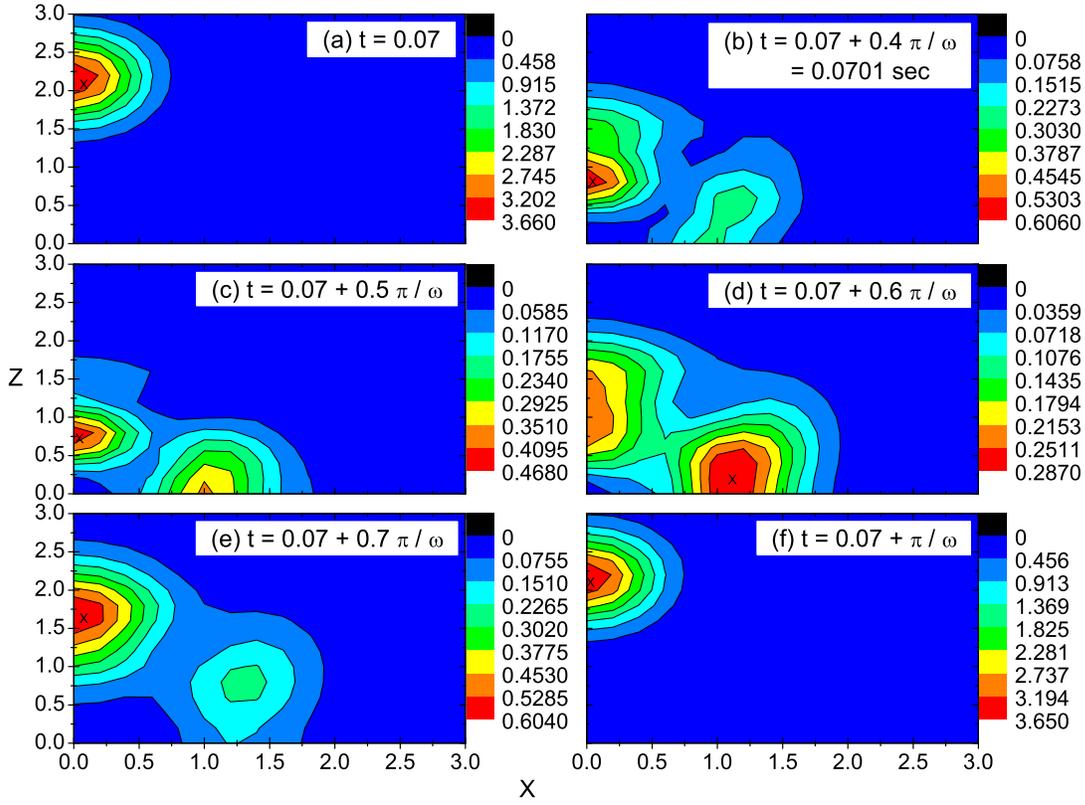}}
 \caption{(Color online) The same as Fig.~\ref{revfig1} but $t$ is given in the
interval ($0.07$ to $0.07+\pi/\omega$).} \label{revfig2}
\end{figure}

The number of collisions that the two atoms have experienced within
$t$ is $\sim t\omega/\pi$. When $t=0.07\sec$ and
$\omega=2000\times2\pi$, this number is $\approx 280$. When $t$ is
close to $0.07\sec$, $\rho$ is shown in Fig.~\ref{revfig2}. At the
first glance, \ref{revfig2}a is similar to \ref{revfig1}a. However,
the peak in \ref{revfig2}a is considerably lower than that of
\ref{revfig1}a implying that the atoms are not well localized as in
the beginning. In fact, $\rho$ has spread widely in \ref{revfig2}a
and contains a smooth peak at $x=1.6$ and $z=1.4$ (this smooth peak
is too low to be seen in the figure). The two atoms are closer to
each other in \ref{revfig2}b than in \ref{revfig1}b. The density
varies more vigorously along $\theta$ in \ref{revfig2}c than in
\ref{revfig1}c. \ref{revfig2}b and \ref{revfig2}d do not have the
approximate similarity as shown previously in \ref{revfig1}b and
\ref{revfig1}d. From \ref{revfig2}a and \ref{revfig2}f, we know that
a round of collision has been completed in the interval from $0.07$
to $0.07+\pi/\omega$. The evolution in this round is explicitly
different from that in the first round due to the accumulated effect
of $V_{12}$.

The accuracy of the above numerical results depends on the number of
basis functions, which is determined by the scopes of $n$ (from 0 to
$n_{\max}$), $l$ (from 0 to $l_{\max}$), and $N$ (from 0 to
$N_{\max}$). When $n_{\max}=N_{\max}=12$ and $l_{\max}=14$, the
associated results are found to be nearly the same as those by using
$n_{\max}=N_{\max}=10$ and $l_{\max}=12$. Thus we believe that the
former choice is sufficient.

\section{Time-dependent probability of the spin-component and the evaluation}

Due to the spin-dependent interaction, spin-flips will occur during
the evolution. From Eq.~(\ref{e06_Psit}) the time-dependent
probability of the spin-component of an atom in $\mu$ is
\begin{eqnarray}
 P_{\mu}(t)
 &=&\sum_{Ji\gamma ,J'i'\gamma'}
    \delta_{n'n}
    \delta_{l'l}
    \cos[(E_{i'}^{J'}-E_i^J)\tau]  \nonumber \\
 &&\cdot
    b_{i'}^{J'}
    C_{i'\gamma'}^{J'}
    b_i^J
    C_{i\gamma}^J
    \sum_{\lambda}
    (2\lambda+1)
    \sqrt{(2S'+1)(2S+1)}  \nonumber \\
 &&\cdot
    W(l3J'3;\lambda S')
    W(l3J3;\lambda S)  \nonumber \\
 &&\cdot
    C_{\lambda ,-\mu ;\ 3\mu }^{J',0}
    C_{\lambda ,-\mu ;\ 3\mu }^{J,0},
 \label{e09_Pmut}
\end{eqnarray}
where both the Clebsch-Gordan and Wigner coefficients
\cite{r_EAR1957} have been introduced. To show the convergency of
numerical calculation, the dependence of $P_{\mu}(t)$ on the number
of basis functions is given in Tab.~\ref{revtab1}.
\begin{table}[tbph]
\caption{\label{revtab1}$P_3(t)$ at three values of $t$ (in $\sec$).
The parameters are given as $\omega=1000\times 2\pi$, $\alpha=3$,
$|\mathbf{a}|=2$, and $g_0=-g_6/2$. The number of basis functions
depends on $n_{\max}$ and $l_{\max}$ ($N_{\max}=n_{\max}$ is
assumed) listed in the first row.}
 \begin{indented}
 \item[]
 \begin{tabular}{@{}lrrrrrrrrrrrr}
 \br
 $n_{\max}$ and $l_{\max}$ & 10 and 12 & 12 and 14 & 14 and 16 \\
 \mr
 $P_3(0.02)$               & 0.4908    & 0.4914    & 0.4914    \\
 $P_3(0.06)$               & 0.4728    & 0.4733    & 0.4734    \\
 $P_3(0.10)$               & 0.4751    & 0.4758    & 0.4760    \\
 \br
 \end{tabular}
 \end{indented}
\end{table}

It is clear from the table that, when $n_{\max}=N_{\max}=12$ and
$l_{\max}=14$ as we have adopted, qualitatively accurate results can
be obtained.
\begin{figure}[htbp]
 \centering
 \resizebox{0.95\columnwidth}{!}{\includegraphics{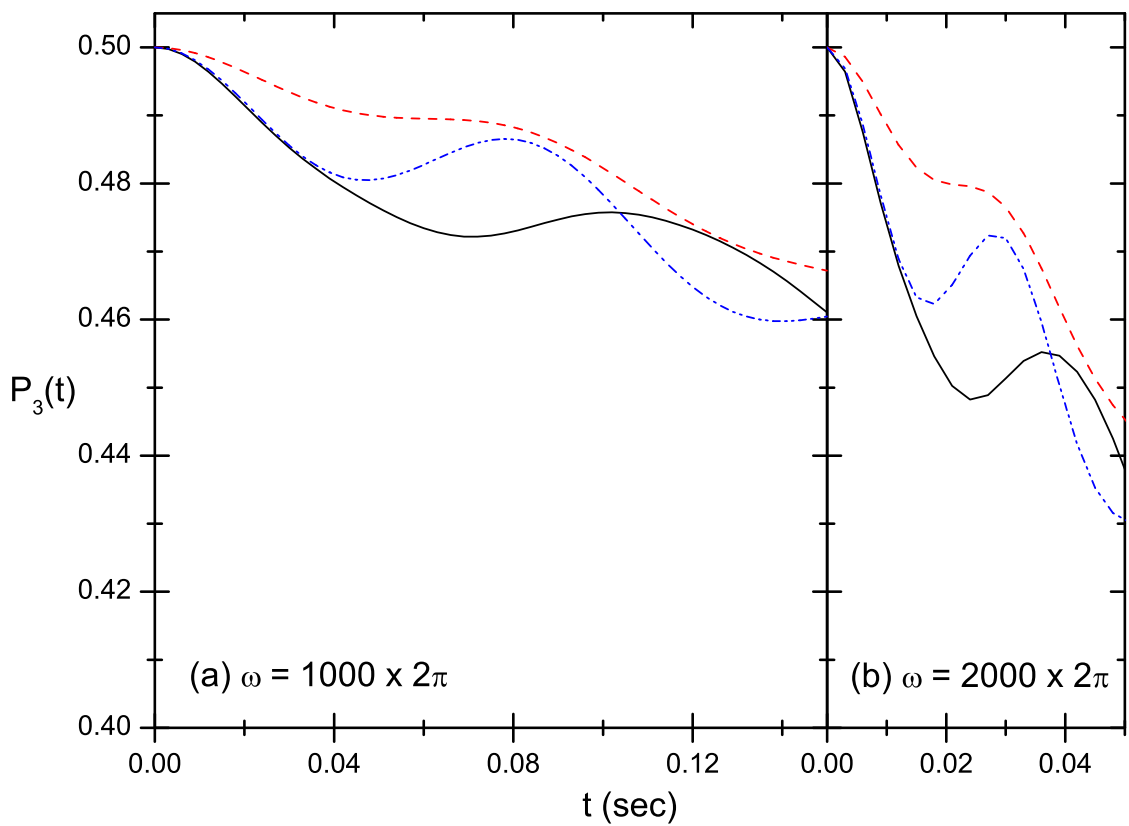}}
 \caption{(Color online) The evolution of $P_3(t)$. The solid, dash,
and dash-dot-dot curves have $g_0=-g_6/2$, $0$, and $g_6/2$,
respectively. In \ref{revfig3}a, $\omega=1000\times 2\pi$,
$\alpha=3$, $|\mathbf{a}|=2$, while in \ref{revfig3}b,
$\omega=2000\times 2\pi$, $\alpha=1.5$, $|\mathbf{a}|=2$.}
\label{revfig3}
\end{figure}

\begin{figure}[htbp]
 \centering
 \resizebox{0.95\columnwidth}{!}{\includegraphics{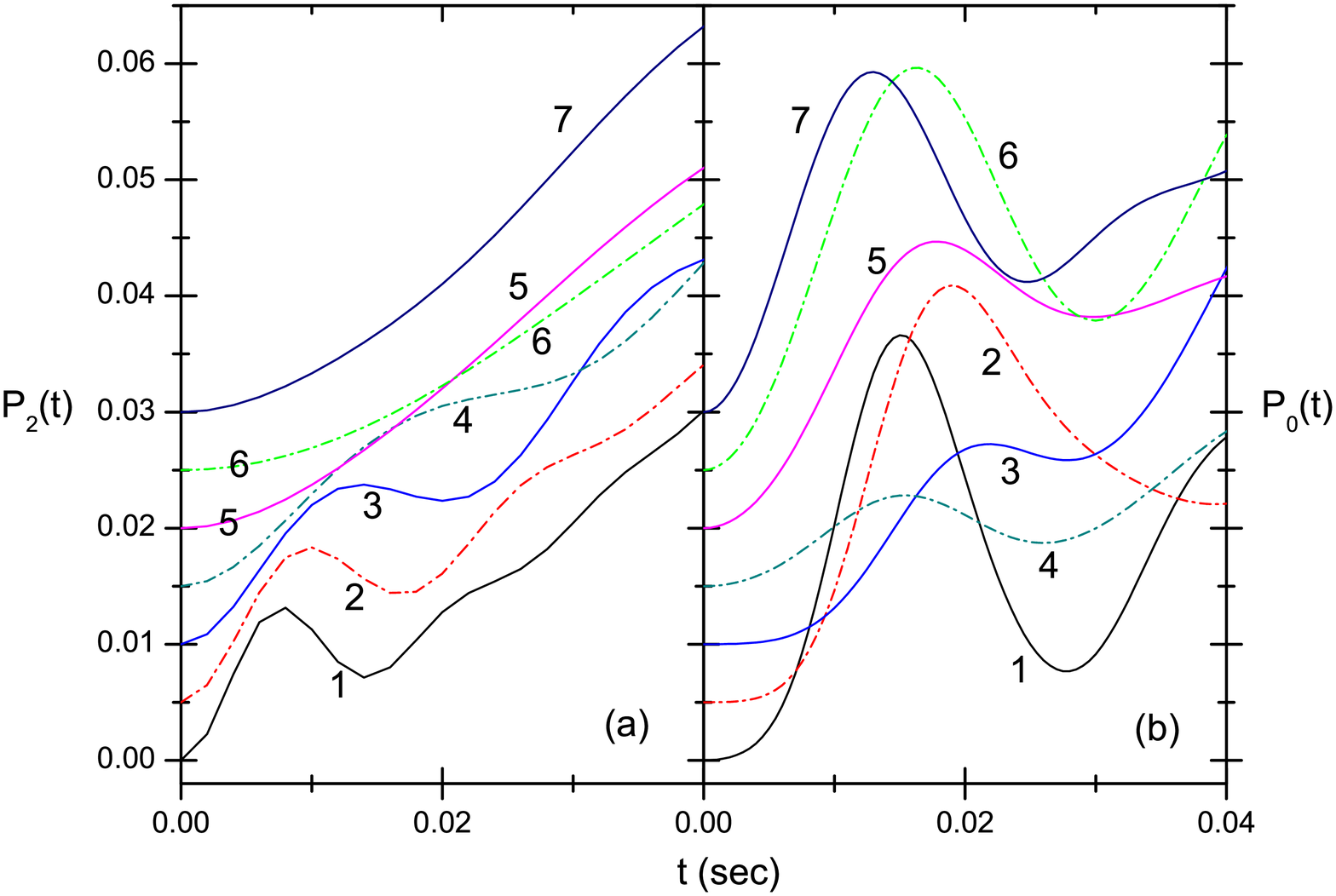}}
 \caption{(Color online) The evolution of $P_2(t)$ (a) and $P_0(t)$
(b). The curves from "1" to "7" have $g_0$ from $-3g_6/4$ to
$3g_6/4$, respectively, with an increase $g_6/4$ in each step. Each
curve has shifted up by $0.005$ more than its adjacent lower
neighbor to guide the eyes. The other parameters are the same as in
Fig.~\ref{revfig1}.} \label{revfig4}
\end{figure}

Due to the symmetry inherent in $\Psi_I$ and in $H_{evol}$,
$P_{-\mu}(t)=P_{\mu}(t)$. The variation of $P_3(t)$ against $t$ is
shown in Fig.~\ref{revfig3}, where $g_0$ is given at three values.
It is recalled that the first round of collision ends at
$t=\pi/\omega=0.00025\sec$. In this very short time, the change of
$P_3(t)$ is negligible. However, after hundreds of repeating
collisions, the dependence of $P_3(t)$ on $g_0$ can be clearly
detected as shown in the figure. Therefore, $P_{\mu}(t)$ can be used
to evaluate $g_0$. In particular, the dependence will become more
explicit if $\omega$ is larger (comparing \ref{revfig3}b with
\ref{revfig3}a). The evolutions of $P_2(t)$ and $P_0(t)$ are shown
in Fig.~\ref{revfig4}a and \ref{revfig4}b, respectively, where $g_0$
is given at seven values. One can see that there is a small peak in
\ref{revfig4}a appearing in the early stage of evolution if $g_0$ is
negative. Its height depends on how negative $g_0$ is, and it would
disappear if $g_0$ is positive. Therefore, the existence or not of
this peak can be used to judge the sign of $g_0$. Together with the
height and location of the second peak in Fig.~\ref{revfig3} (or the
first peak of \ref{revfig4}b), the strength $g_0$ can be known.

Since $V_{dd}$ can alter $l$, the spatial structure can thereby be
altered. On the other hand, $V_{dd}$ can also alter $S$; therefore
spin-evolutions are also affected. We found that, if $V_{dd}$ is
reduced (strengthened), the spin-flips would occur less (more)
probable. E.g., when $V_{dd}$ is changed to $\beta V_{dd}$, the
values of the first minimum of the solid curve of
Fig.~\ref{revfig3}b would be $0.462$, $0.448$, and $0.298$,
respectively, if $\beta=0$, $1$, and $10$. It implies that a strong
dipole force will cause strong spin-flips. This is a notable point.
On the other hand, recent experimental progress suggests that
condensation of molecules with large permanent dipole moments, such
as OH \cite{r_SYT2005,r_JRB2004}, RbCs \cite{r_JMS2005}, KRb
\cite{r_DW2004}, and NH \cite{r_DE2004}, may be achieved. These
systems would have very strong dipole interaction, $10^2$ or more
times stronger than in chromium. Therefore, distinguished phenomena
of spins caused by the very strong $V_{dd}$ are expected.

In order to understand better the phenomenon of spin-flips, the
momentum-spin correlation is studied in the following. Let the
coordinates $\mathbf{r}_1$ and $\mathbf{r}_2$ be introduced via the
Talmi-Moshinsky (T-M) coefficients as
\begin{eqnarray}
 &&
 \bar{\phi}_{N,0}(R)
 \phi_{nl}(r)
 Y_{00}(\hat{R})
 Y_{lm}(\hat{r}) \nonumber \\
 &=&
  \sum_{n_1 l_1 n_2 l_2}
  a_{n_1 l_1 n_2 l_2}^{N0nl,l}
  \hat{\phi}_{n_1 l_1}(r_1)
  \hat{\phi}_{n_2 l_2}(r_2)
  [Y_{l_1}(\hat{r}_1)\ Y_{l_2}(\hat{r}_2)]_{lm},
 \label{e10_phiphiYY}
\end{eqnarray}
where $l_1$ and $l_2$ are coupled to $l$ and $m$, and
$\hat{\phi}_{nl}(r_i)Y_{lm_l}(\hat{r_i})$ are the eigenstates of
$H_{r_i}\equiv -\frac{1}{2}\nabla _{r_i}^2+\frac{1}{2}r_i^2$. The
analytical expression of T-M coefficients has been given in the
refs. \cite{r_WT1981,r_BM1966,r_BTA1960}. Inserting
Eq.~(\ref{e10_phiphiYY}) into (\ref{e06_Psit}), we have
\begin{eqnarray}
 \Psi (\tau)
 &=&\sum_{n_1 l_1, n_2 l_2, m_1 m_2, \mu\nu}
    \delta_{m_1 + m_2,\ -\mu -\nu}
    Y_{n_1 l_1, n_2 l_2,m_1 m_2}^{\mu\nu}(\tau) \nonumber \\
 &&\cdot
    \hat{\phi}_{n_1 l_1}(r_1)
    \hat{\phi}_{n_2 l_2}(r_2)
    Y_{l_1 m_1}(\hat{r}_1)
    Y_{l_2 m_2}(\hat{r}_2) \nonumber \\
 &&\cdot
    \chi_{\mu}(1)\chi_{\nu}(2),
 \label{e11_Psitau}
\end{eqnarray}
where $Y_{n_1 l_1, n_2 l_2, m_1 m_2}^{\mu\nu}(\tau)$ can be known by
comparing Eq.~(\ref{e11_Psitau}) with (\ref{e06_Psit}) and
(\ref{e10_phiphiYY}).

Insert Eq.~(\ref{e11_Psitau}) into the normality
$\langle\Psi(\tau)|\Psi(\tau)\rangle=1$, it is straight forward to
obtain the time-dependent probability of finding a particle in
$\chi_{\nu}$ state at $\mathbf{r}_2$ and $\tau$ as follows
\begin{eqnarray}
 P_{\nu}(\mathbf{r}_2,\tau)
 &=&\sum_{n_2' l_2', n_2 l_2, m_2}
    Z_{n_2' l_2', n_2 l_2, m_2}^{\nu}(\tau)
    \hat{\phi}_{n_2' l_2'}(r_2)
    \hat{\phi}_{n_2 l_2}(r_2) \nonumber \\
 &&\cdot
    Y_{l_2' m_2}^*(\hat{r}_2)
    Y_{l_2 m_2}(\hat{r}_2),
 \label{e12_Pnu}
\end{eqnarray}
where
\begin{eqnarray}
 Z_{n_2' l_2', n_2 l_2, m_2}^{\nu}(\tau)
 &=&
   \sum_{n_1 l_1 m_1 \mu}
   \delta_{m_1+m_2,\ -\mu -\nu}
   [Y_{n_1 l_1, n_2' l_2', m_1 m_2}^{\mu\nu}(\tau)]^* \nonumber \\
 &&\cdot
   Y_{n_1 l_1, n_2 l_2, m_1 m_2}^{\mu\nu}(\tau).
 \label{e13_Z}
\end{eqnarray}
This probability fulfills $\sum_{\nu}\int d\mathbf{r}_2\
P_{\nu}(\mathbf{r}_2,\tau)=1$.

For observing the momentum-spin correlation, we define
\begin{eqnarray}
 \xi_{nl}(p)
 =\sqrt{\frac{2}{\pi}}
  (-i)^l\int r^2dr\
  j_l(pr)
  \hat{\phi}_{nl}(r),
 \label{e14_xinlp}
\end{eqnarray}
where the spherical Bessel function has been introduced. Then from
Eq.~(\ref{e12_Pnu}), we obtain
\begin{eqnarray}
 Q_{\nu}(\mathbf{p},\tau)
 &=&
  \sum_{n_2' l_2', n_2 l_2, m_2}
  Z_{n_2' l_2', n_2 l_2, m_2}^{\nu}(\tau)
  \xi_{n_2' l_2'}^*(p)
  \xi_{n_2 l_2}(p) \nonumber \\
 &&\cdot
  Y_{l_2' m_2}^*(\hat{p})
  Y_{l_2 m_2}(\hat{p}),
  \label{e15_Qnu}
\end{eqnarray}
which is the probability of finding a particle in $\nu$ at
$\mathbf{p}$ and $\tau$. It fulfills $\sum_{\nu}\int d\mathbf{p}\
Q_{\nu}(\mathbf{p},\tau)=1$.

In order to observe $Q_{\nu}$ at a given time $\tau_c$, the broad
trap has to be suddenly cancelled at $\tau_c$. Then, the particles
begin to go out and the successive evolution is an expansion and is
governed by the Hamiltonian $H_{free}+V_{12}$, where
$H_{free}=-\frac{1}{2}(\nabla_1^2+\nabla_2^2)$. During the
expansion, collisions happen scarcely. We have already seen that the
effect of $V_{12}$ within a few collisions is negligible. Therefore,
it is safe to neglect $V_{12}$ during the expansion, thus the wave
function after $\tau_c$ is $\Psi(\tau>\tau_c) \approx
e^{-iH_{free}(\tau-\tau_c)} \Psi(\tau_c) \equiv \Psi_{free}(\tau)$.

It is straight forward to see that the weight of $\Psi_{free}(\tau)$
in $|\mathbf{p}_1, \mathbf{p}_2 \rangle$ is the same as that of
$\Psi(\tau_c)$. Therefore, after the cancellation, the probability
$Q_{\nu}$ depends only on $\tau_c$ but not on $\tau$. Thus, the time
of observation $\tau$ is not relevant, but the choice of $\tau_c$ is
essential.

\begin{figure}[htbp]
 \centering
 \resizebox{0.95\columnwidth}{!}{\includegraphics{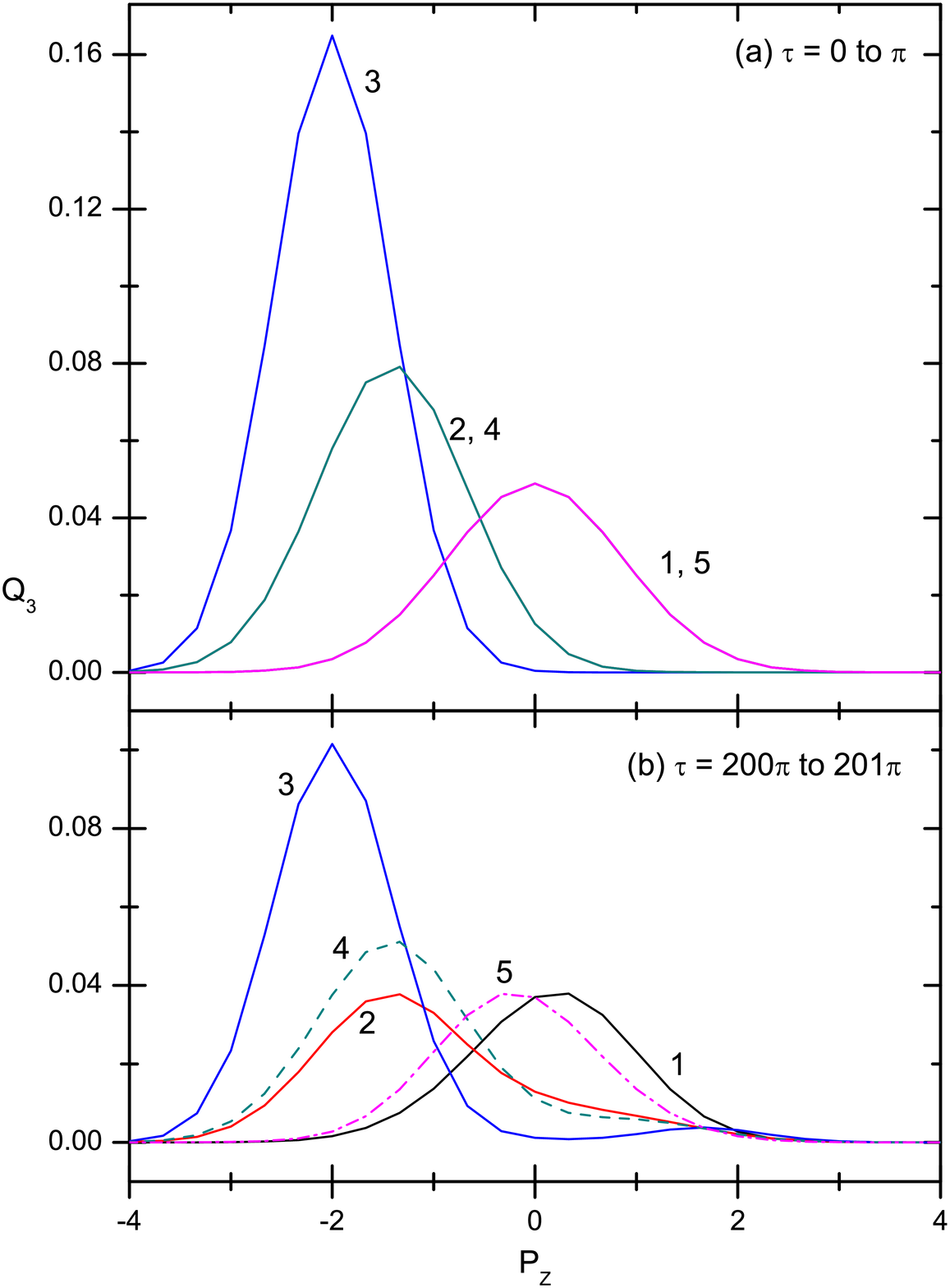}}
 \caption{(Color online) The distribution of $Q_3(\mathbf{p},\tau_c)$
over the $Z$-component of $\mathbf{p}$. $\tau_c$ is the time to
cancel the trap, and is given at five values as $\tau_c=(i-1)\pi
/4+\tau_X$, "$i$" is marked by the associated curve, and $\tau_X=0$
(a) and $200\pi$ (b). The unit of momentum in this paper is
$\sqrt{m\hbar\omega}$. The parameters are the same as in
Fig.~\ref{revfig1}.} \label{revfig5}
\end{figure}

An example of $Q_3(\mathbf{p},\tau_c)$ in the earliest stage of
evolution is given in Fig.~\ref{revfig5}a. Curve "1" describes the
momentum distribution of the initial h.o. ground state over the
$Z$-component of $\mathbf{p}$. When the evolution begins, the
distribution shifts to the left as shown by "2". It implies that the
component with $\nu =3$ is moving from the initial end toward the
center. When $\tau =\pi /2$, a sharp peak appears at $p_{z}=-2$ as
shown by "3". Meanwhile, as we know from Fig.~\ref{revfig1}c, the
particle is close to the center, and it has a maximal momentum
pointing to the opposite end. Afterward, the magnitude of momentum
begins to decrease. When $\tau =\pi$, the average momentum is again
zero as shown by "5" which overlaps "1". Then the process proceeds
in reverse direction, and repeatedly. When $\tau$ is larger, the
periodicity will be gradually spoiled by the interaction. This is
shown in Fig.~\ref{revfig5}b, where "5" overlaps "1" no more.
Instead, the peak of "5" shifts a little to the left. It implies
that the peak has not yet completely arrived at the opposite end,
its arrival is delayed.

The momentum-spin correlation can be revealed by
$Q_{\nu}(\mathbf{p},\tau_c)$. However, we are more interested in a
quantity which is easier to be observed. Therefore, we define
$Q_{\nu}^I(\theta_p,\tau_c) \equiv \int p^2\ dp\
Q_{\nu}(\mathbf{p},\tau_c)$, where $\theta_p$ is the angle between
$\mathbf{p}$ and the $+Z$-axis (the azimuthal angle of $\mathbf{p}$
is irrelevant). This is the probability of finding a particle in
$\nu$ and emitting at a specified direction $\theta_p$ at $\tau_c$.
It fulfills $2\pi \int \sin \theta_p\ d\theta_p\
Q_{\nu}^I(\theta_p,\tau_c)=1$. This quantity is much easier to be
experimentally measured because only the orientation of spin is
concerned, which can be measured via a Stern-Gerlach device setting
towards the outgoing angle $\theta_p$.

\begin{figure}[htbp]
 \centering
 \resizebox{0.95\columnwidth}{!}{\includegraphics{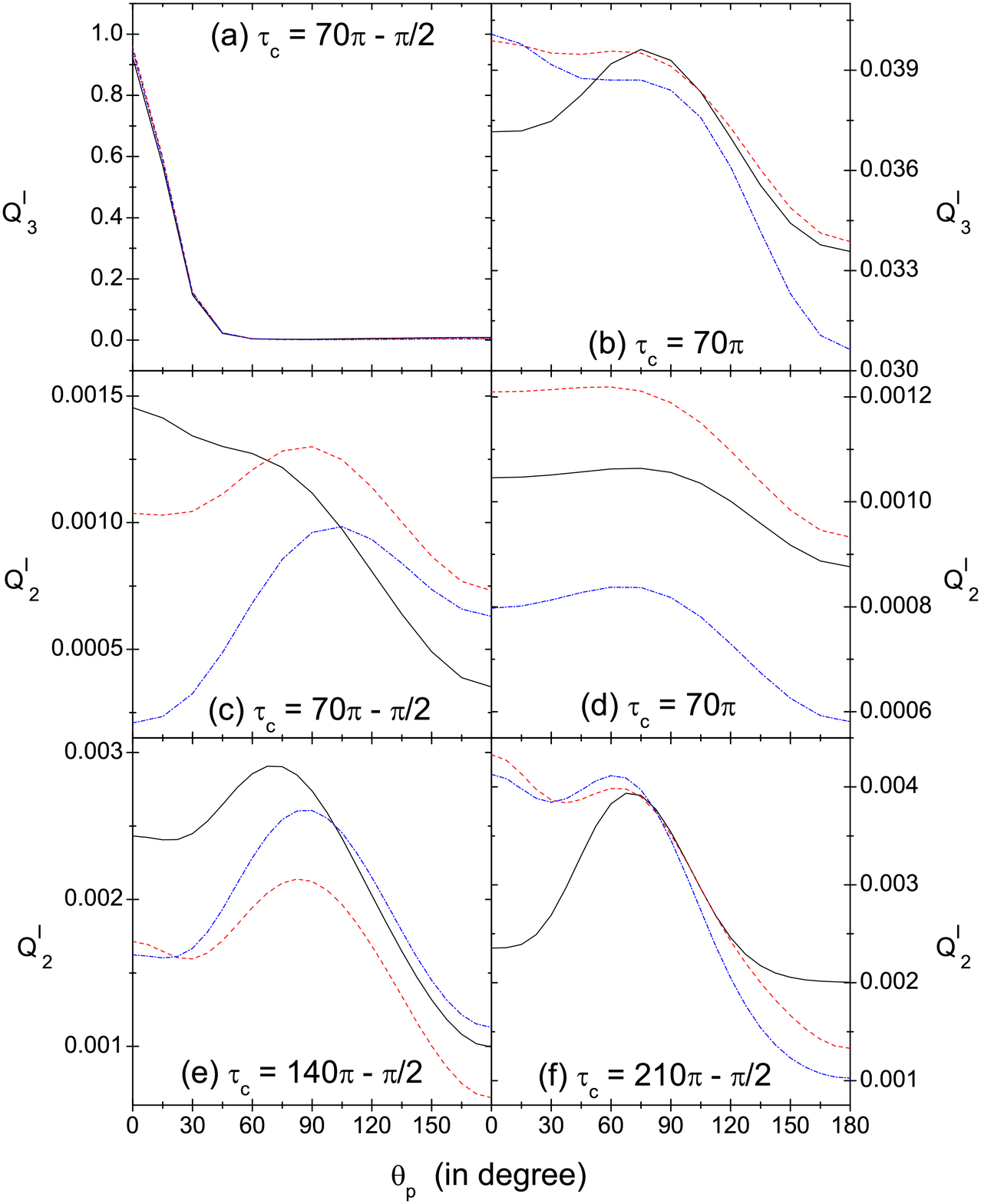}}
 \caption{(Color online) Angular distribution of
$Q_{\nu}^I(\theta_p,\tau_c)$ against $\theta_p$. $\nu=3$ for (a) and
(b), and $\nu=2$ for the others. $\tau_c$ is fixed and marked on the
panels. $g_0$ is given at three values $-g_6/4$, $0$ and $g_6/4$,
respectively, and the associated curves are in solid, dash, and
dash-dot lines. The other parameters are the same as in
Fig.~\ref{revfig1}.} \label{revfig6}
\end{figure}

$Q_{\nu}^I$ is sensitive to $\tau_c$ (therefore both the time of
creation and the time of cancellation of the broad potential should
be precise). For examples, when $\tau_c=69.5\pi$ (meanwhile the
average positions of both particles are close to the center),
$Q_3^I$ against $\theta_p$ with $g_0$ given at three presumed values
are plotted in Fig.~\ref{revfig6}a, where the momentum is
distributed around $\theta_p=0$. Thus the particle is rushing back
to the initial end disregarding how $g_0$ is (meanwhile the $\nu=-3$
particle is rushing to the opposite end). If $\tau_c=70.5\pi$, the
peak of \ref{revfig6}a would appear at the right (not yet plotted).
When $\tau_c=70\pi$ (meanwhile the particle is close to the initial
end), $Q_3^I$ depends slightly on $\theta_p$ as shown in
\ref{revfig6}b, the dependence is not very sensitive to $g_0$.

On the other hand, due to the spin-flips, the $|\nu|\neq 3$
components are created and the creation is sensitive to the
interaction. For the case of the $\nu=2$ component plotted in
\ref{revfig6}c, the emission with a smaller $\theta_p$ would be more
probable if $g_0$ is negative, but less probable if $g_0$ is
positive. The probability $Q_2^I(0,\ 69.5\pi)$ would decrease by 7
times if $g_0$ is changed from $-g_6/4$ to $g_6/4$. Thus,
information on the interaction can be extracted.

When $\tau_c$ increases, $Q_2^I$ would also increase gradually. This
is shown in \ref{revfig6}c, \ref{revfig6}e, and \ref{revfig6}f,
where $\tau_c=2k\pi-\pi/2$ (note that the ordinates of these figures
have different scales). An example with $\tau_c=2k\pi$ is shown in
\ref{revfig6}d. Comparing \ref{revfig6}d with \ref{revfig6}c, one
can see how $Q_2^I$ is changed after the interval $\pi/2$.

The features of other $|\nu |\neq 3$ components are more or less
similar to those of $\nu=2$. Information on the interaction can also
be extracted from them. All the curves in Fig.~\ref{revfig6} will
undergo a left-right reflection if $\tau_c$ is changed from
$2k\pi-\pi/2$ to $2k\pi+\pi/2$, or if $\nu$ is changed to $-\nu$.
The reflection is nearly exact.

It is emphasized that the dependence on interaction is hardly found
in the early stage. For an example, when $\tau_c=2\pi$ (meanwhile
both particles nearly completely return to their initial positions
after two rounds of collisions), the curves of \ref{revfig6}b will
become a set of horizontal lines and very close to each other, and
therefore no valuable information can be extracted. Thus the
introduction of the trap, that leads to repeated collisions, is
necessary so that the weak effect of interaction can be accumulated.

\begin{figure}[htbp]
 \centering
 \resizebox{0.95\columnwidth}{!}{\includegraphics{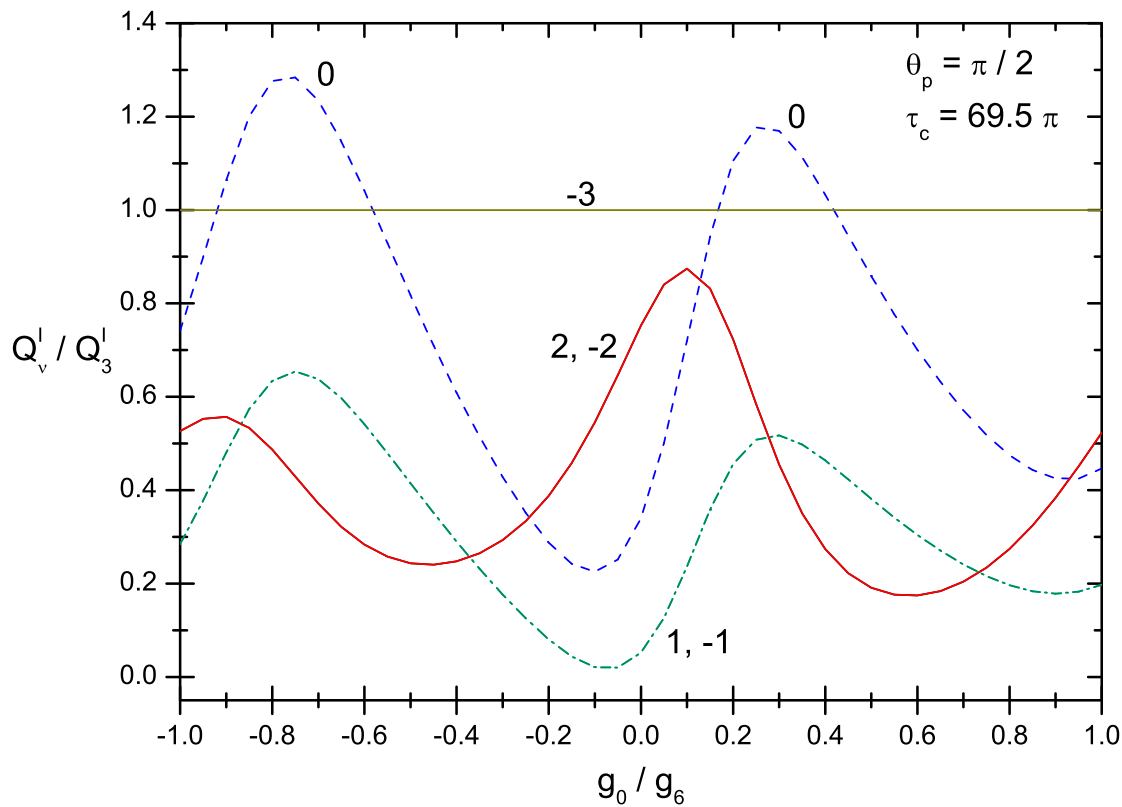}}
\caption{(Color online) $Q_{\nu}^I/Q_3^I$ with $\theta_p=\pi/2$ and
$\tau_c=69.5\pi$ against $g_0/g_6$. $\nu$ is marked by the curves.
The other parameters are the same as in Fig.~\ref{revfig1}.}
\label{revfig7}
\end{figure}

When $\theta_p$ and $\tau_c$ are both fixed, an example is shown in
Fig.~\ref{revfig7} where the ratio $Q_{\nu}^I/Q_3^I$ with
$\theta_p=\pi/2$ and $\tau_c=69.5\pi$ is plotted against $g_0/g_6$.
The curves with $\nu$ and $-\nu$ overlap with each other entirely
(however, if $\theta_p\neq \pi/2$, they do not). It is clear from
the figure that $g_0$ can be uniquely determined if a few ratios can
be measured (say, those with $\nu=2$ and 0).

\section{Conclusions}

In summary, an idea is proposed to study the trapped 2-body
scattering. The two particles are first localized, then they collide
with each other repeatedly in a trap, then they escape. Related
theoretical derivation and numerical calculation have been performed
to study the coordinate-spin and momentum-spin correlations. The
$^{52}$Cr atoms have been chosen as an example. Due to the repeating
collisions in the trap, the effect of the weak interaction can be
accumulated and enlarged, and the strength $g_0$ can be thereby
determined. It is expected that the approach might open a new way
for studying various spin-dependent interactions among atoms
(molecules) with nonzero spin, in particular for the studies of very
weak forces.

\ack

The support from the NSFC under the grant 10874249 and from the
project of National Basic Research Program of China (2007CB935500)
is appreciated.

\appendix

\section{Matrix elements of $V_{12}$ between the basis functions}

\begin{eqnarray}
 &&
 \langle
 \phi _{n'l'}(r)
 (l'S')_{J'}
 |V_{12}|
 \phi _{nl}(r)(lS)_{J}
 \rangle \nonumber \\
 &=&\delta _{J',J}
    \Big[
     \frac{1}{4\pi }
     g_{S}
     \delta _{l'0}
     \delta _{l0}
     \delta_{S'S}
     \phi _{n'0}(0)
     \phi _{n0}(0)
    -252\sqrt{5}C_{d}
     \sqrt{(2S'+1)(2S+1)(2l+1)} \nonumber \\
 & &\cdot
    C_{1,0,\ 1,0}^{2,0}
    C_{l,0,\ 2,0}^{l',0}
    W(l2jS';l'S)
    \left\{
    \begin{array}{ccc}
     1 & 1 & 2 \\
     3 & 3 & S \\
     3 & 3 & S'
    \end{array}
    \right\}
    \int \frac{dr}{r}\phi _{n'l'}(r)\phi _{nl}(r) \Big],
\end{eqnarray}
where the Clebsch-Gordan, Wigner and 9-$j$ symbols \cite{r_EAR1957}
are introduced. Furthermore, due to the Boson statistics, both $s+l$
and $s'+l'$ should be even.

\end{document}